\documentclass[paper,ras]{agutex2015}
\usepackage{graphicx}

\authorrunninghead{CORT\'ES, REEVES \& BUSTOS}

\titlerunninghead{Distribution of PWV in the Chajnantor Area}

\begin{document}

%
%

\title{Analysis of the distribution of precipitable water vapor in the Chajnantor area.}
%
%

%
%



\authors{Fernando Cort\'es,\altaffilmark{1}
 Rodrigo Reeves,\altaffilmark{1} Ricardo Bustos\altaffilmark{2}}

\altaffiltext{1}{CePIA, Departamento de Astronom\'ia, Universidad de Concepci\'on, Casilla 160 C, Concepci\'on, Chile}

\altaffiltext{2}{Laboratorio de Astro-Ingenier\'ia y Microondas, Facultad de Ingenier\'ia, Universidad Cat\'olica de la Sant\'isima Concepci\'on, Alonso de Ribera 2850, Concepci\'on, Chile}






%
%


\keypoints{ \item We validate a method to convert from sub-mm opacity to PWV.
\item Tipper data was successfully calibrated and correlate to high degree with data from independent sources.
\item Found a PWV ratio between Cerro Chajnantor and Chajnantor Plateau of $0.68$.
}


%
%


\begin{abstract}
In this work, we present results from a long-term precipitable water vapor (PWV) study in the Chajnantor area, in northern Chile. Data from several instruments located at relevant sites for sub-millimeter and mid-infrared astronomy were processed to obtain relations between the atmospheric conditions among the sites. The data used for this study can be considered the richest dataset to date, because of the geographical sampling of the region, including sites at different altitudes, a time span from 2005 to 2014, and the different techniques and instruments used for the measurements. We validate a method to convert atmospheric opacity from 350 $\mu$m tipper radiometers to PWV. An average of 0.68 PWV ratio between Cerro Chajnantor and Llano of Chajnantor was found.
\end{abstract}

%
%

%

\begin{article}

%
%

\section{Introduction}

\hspace{4mm} The Chajnantor area is considered one of the best sites in the world for millimeter, submillimeter and mid-infrared astronomy due to the combination of dryness, i.e. low precipitable water vapor (PWV) and the notable high altitude above $5000$ meters. The area has been selected as the location for world-class  observatories, such as ALMA (\cite{wotten09}), APEX (\cite{gusten06}), CBI (\cite{padin02}), TAO (\cite{motohara11}), and others. A study of the tropospheric distribution of PWV over the area is of relevance to determine the optimum altitude for new observatories that can impact project costs significantly. Another reason to support such a study is to observe temporal drifts and cycles in the atmospheric conditions for the site, as well as contributing with the understanding of calibrations and cross-comparisons between instruments located in the area.

There are a number of publications in the literature referring to PWV and weather conditions in the Chajnantor area and their implications to astronomy. For example, in \cite{delgado99}, several empirical relations for atmospheric variables are derived to understand the climatology in the Chajnantor area and how to obtain the amount of PWV from radiometric measurements. Long-term studies using climatological data for the Chajnantor area was presented by \cite{bustos00} and \cite{otarola05}. The phase correction by using the PWV at the line of sight in interferometric mode at ALMA was studied by \cite{nikolic13}. PWV estimates and forecasts for the Chajnantor area using GOES satellites were produced by \cite{marin15}. An atmospheric measurement campaign was performed by \cite{turner10} on Cerro Toco in 2009. The atmospheric transparency has been intensively studied by \cite{radford00}, \cite{giovanelli01}, \cite{radford11}, and \cite{radford16}. In addition, the differences in PWV between the Chajnantor Plateau and Cerro Chajnantor was shown in \cite{bustos14} for a time span of 5 days. 

The amount of PWV obtained by each instrument located in the Chajnantor area is measured independently and is only valid for that location and time. There are no studies involving the spatial distribution of PWV over the Chajnantor area and its temporal variations, nor long term relations that connect the amount of PWV at  different locations and measured with different instruments.

In this work, we have derived a new method to convert atmospheric opacity (optical depth) from $350 \,\mu m$ tipper radiometers to PWV, using the AM spectroscopy/atmospheric model (\cite{paine16}). We have aimed this study to understand the ratio of PWV between the Chajnantor Plateau (5080\,m of altitude), Cerro Toco (5320\,m), and the summit of Cerro Chajnantor (5612\,m).

\subsection{Instruments}

\hspace{4mm} Different instruments have been located on the Chajnantor area over the years. These instruments differ in their technical characteristics and specifications, operations approach, location, altitude, and also in the observables they measure. The instruments used in this study, their locations, and time span are detailed in Table \ref{tab-data-inst}.

Water vapor radiometers (WVR) provide a measure of the atmospheric brightness temperature of a selected water molecule vibro-rotational absorption/emission band. Then, an atmospheric model is used to compute PWV.  This technique is used by APEX, RHUBC-II and UdeC, as shown in Table \ref{tab-data-inst}.

The APEX telescope includes a 183 GHz water vapor radiometer installed in the Cassegrain cabin. The instrument measures PWV whenever the telescope shutter is open. We have decided to use the data from this instrument as the main reference for comparisons among the selected sites.

On the other hand, the PWV at mid-infrared was estimated from atmospheric extinction using photometry, as for the TAO data. The submillimeter tippers measure sky brightness temperature at several zenith angles and fit for optical depth, at a certain bandwidth defined by the optical filter at the input of the instrument. 

\subsection{Software}
The computational tools used to produce all graphs, as well as derive the physical relationships and results include: AM 9.0 (Atmospheric Model, \cite{paine16}), Python version 2.7.2, and TOPCAT (Tool for OPerations on Catalogues And Tables) version 4.1.

\begin{table*}[HB!]
\caption{Instruments used in this study and their location, altitude, observable and time span. }
\centering
\begin{tabular}{|c |c | c | c | c | c | c |}
\hline

Instrument & Location    &    Altitude (m)  &      Observable  & Time span & ID \\
                  &                   &          &                              &                 &  \\ \hline \hline
 APEX         &  Chajnantor  &  5107   &   PWV   & 2006 to 2014  & APEX \\ 
radiometer        &  plateau  &          &  from WVR at 183 GHz&   & \\  \hline  
TAO            & Cerro Chajnantor    &  5640   & PWV  from  mid-infrared  & 2009 to 2011  & TAO \\ 
                   & summit  &     &  astronomical  observations & & \\  \hline 

RHUBC-II   &  Cerro   & 5320   & PWV 	 from WVR    &   August to   October & TOCO \\ 
                   &  Toco  &     & at 183 GHz    &  2009    & \\    \hline

       &  Chajnantor  &  5080   & Opacity    & 1997 to 2005  &  TA-1 \\ 
   &  Plateau (NRAO)&     &    at 350 $\mu m$ &   & \\ 
                        &   & & & & \\
                     
 Tipper                         & Chajnantor  & 5080    & Opacity       &   2005 to 2010  &TA-2 \\ 
 radiometer A                          &  plateau  (CBI)&     & at 350 $\mu m$    &     & \\ 
                         &     &     &     &     & \\ 

                         & Chajnantor  &  5107   & Opacity       &   2011 to now &  TA-3\\ 
                         &  plateau  (APEX)  &     &   at 350 $\mu m$   &    &  \\   \hline

         &  Chajnantor  &  5080   & Opacity    & 2000 to 2005  & TB-1  \\ 
    &  Plateau (NRAO) &     & at 350 $\mu m$    &    & \\ 
                         &     &     &     &     & \\ 

Tipper                           & Chajnantor  & 5080    & Opacity      &   2005 to 2009  & TB-2 \\ 
 radiometer B                         &  plateau (CBI)  &     &  at 350 $\mu m$    &   &  \\ 
                                                  &     &     &     &     & \\ 

                         &  Cerro &  5612   & Opacity        &   2009 to now  &TB-3 \\ 
                         & Chajnantor   &     &  at 350 $\mu m$  &      & \\   
                                       & summit     &     &     &     & \\  \hline  
               & Cerro           & 5612    &  PWV    &  2011  & UdeC \\   
UDEC                           &  Chajnantor  &      & at 183 GHz   &   & \\ 
                                                                 & summit     &     &     &     & \\  
\hline  
\end{tabular}
\label{tab-data-inst}
\end{table*}

\section{Converting $350\,\mu m$ opacity to preci-
pitable water vapor}

The largest dataset, in time and locations, are supplied by the $350\,\mu m$ tipper radiometers. These instruments deliver atmospheric opacity instead of PWV. For this reason, a method has been developed to convert the measured atmospheric opacity by the tippers into PWV and hence, allow for the comparison of water vapor content and its variability between the different sites under study. The converted PWV from the tippers was calibrated using measurements from APEX as our reference standard.\\


The transformation is modeled by a linear relation between PWV and opacity at 350 $\mu m$. The AM atmospheric modeling software, was used to emulate the functionality of the tippers from mechanical and radiometric perspectives, with the aim of deriving a single  opacity value from an input PWV value. The AM software has been recently updated to version 9.0, which we use throughout the analysis.

\subsection{Input configuration files for AM}
The AM software is a tool for radiative transfer computations at microwave to submillimeter wavelengths for modeling propagation paths. The user may define a configuration file in which the layers of a propagating medium, here the atmosphere, are defined by their main constituents and physical state, i.e. layer average temperature, pressure, water and ozone content, among a number of available variables.

In this work, and following suggestions by AM's author, we generated realistic atmospheric configuration files for the 3 different sites in which tipper radiometers were deployed, i.e. CBI site, APEX site and Cerro Chajnantor, based on NASA MERRA-2 reanalysis data (\cite{molod15}) extracted specifically for the coordinates of the Chajnantor area. The variables taken from the database to build the models were base temperature and pressure for each layer. Average water vapor and ozone volume mixing ratios were derived, for each layer, from the mass mixing ratios available in the MERRA-2 dataset. The time-average surface pressure and temperature used in the models for each site were: $556\,mbar$ and $272.4\,K$ for the CBI site, $554\,mbar$ and $272.3\,K$ for the APEX site, and $518\,mbar$ and $268.6\,K$ for Cerro Chajnantor.

\subsection{Implementation}
\begin{figure}[h]

\noindent\includegraphics[width=20pc]{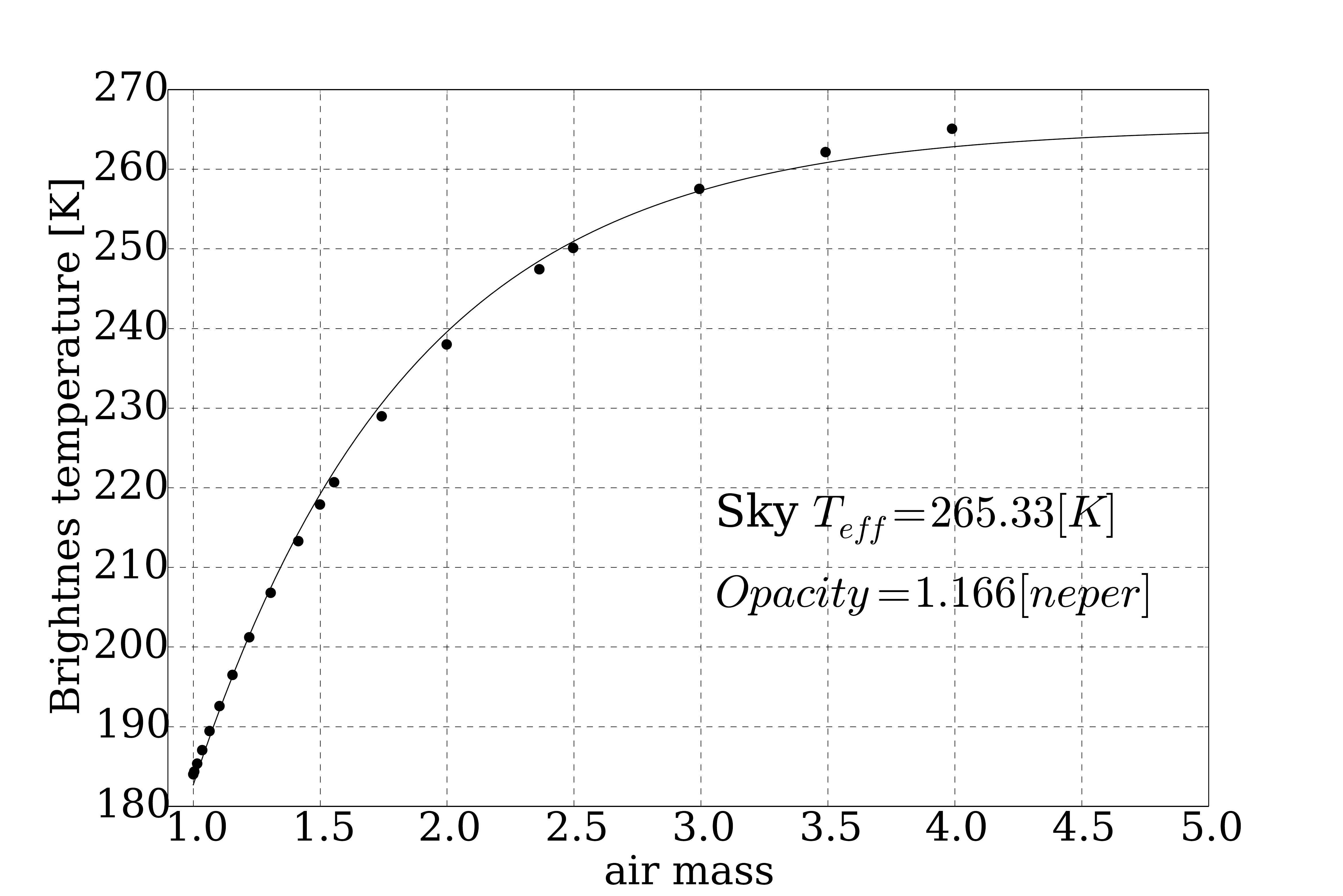}
\caption{Sample fitting procedure for airmass and average sky brightness temperature at Chajnantor Plateau. The PWV input in this case was 0.75 mm.}
\label{figure1}
\end{figure}

The $350\,\mu m$ tipper radiometers estimate the atmospheric opacity measuring the sky brightness temperature at 7 different air masses (1, 1.5, 2, 2.5, 3, 3.5, 4). By using an exponential fit, a single value for the atmospheric opacity is determined (\cite{radford16}). The optics of the radiometer tip from zenith to the highest zenith angle and back to zenith in about 13 minutes. To model the instrument functionality, the spectrum for the sky brightness temperature was obtained with AM 9.0 using a selected airmass and PWV as input parameters to the software run. Then, the effective brightness temperature input to the radiometer was obtained as the average of the sky brightness temperature spectrum over a frequency range defined by the parameters of the tipper input filter, i.e. central frequency of 850 GHz and 103 GHz of bandwidth (\cite{radford16}). Once this was done at each airmass, an exponential fit was used to derive the effective sky temperature and the atmospheric opacity, as shown in Figure~\ref{figure1}. Note that it was decided to use a denser sampling in air masses to facilitate the exponential fitting procedure at extreme input conditions, going from 7 samples to 19 steps in airmass. 

The chosen characteristic function for airmass $(X)$ vs average sky brightness temperature $(\bar {T_{b}})$ is as follows:

\begin{equation}\eqnum{1}
\bar{T_{b}}= A\cdot(1-e^{-B{X}})
\label{eq1}
\end{equation}

\noindent where, $A$ is effective sky radiometric temperature and $B$ is the microwave atmospheric opacity for a given PWV. The function was fitted by non-linear least squares using Levenberg-Marquardt in Python.\\

This procedure is repeated for a list of PWV values. The results for the fits including parameter uncertainties from the fitting algorithm are listed in Table \ref{tab-pwv}. The results in this table are used to make Figure \ref{fig3a}.

\begin{table}[h]
\caption{Results of opacity and sky temperature for all the PWV in Chajnantor Plateau, simulated  with AM.  }
\centering
\begin{tabular}{|c |c | c | c | c |c| }
\hline
 PWV           & Sky               & Opacity  & Error             & Error   \\ 
 mm        & Temperature & (B) [n]    & Temperature &  opacity   \\ 
                    & (A) [K]           &               &                       &  \\ \hline \hline

0.25    &     240.6   &  0.60     &  2.3  &   0.01   \\ \hline         
0.51    &     259.7   &   0.880  &  1.1  &  0.009	   \\ \hline
0.77    &     265.3   &   1.166  &  0.7  & 0.008	   \\ \hline
1.03    &     267.6   &   1.454  &  0.4  & 0.009	   \\ \hline
1.29    &     268.8   &   1.739  &  0.3  &  0.009	   \\ \hline
1.54    &     269.4   &    2.02   &  0.2  & 0.01	   \\ \hline
1.80    &     269.9   &    2.30   &  0.2  & 0.01	   \\ \hline
2.06    &     270.2   &    2.57   &  0.1  & 0.01	   \\ \hline
2.32    &     270.4   &    2.84   &  0.1  & 0.01	   \\ \hline
2.57    &     270.6   &    3.10   &  0.1  & 0.01	   \\ \hline
2.83    &     270.7   &    3.35   &  0.1  &  0.02	   \\ \hline
3.09    &     270.9   &    3.60   &  0.1  & 0.02	   \\ \hline
3.61    &     271.08  &   4.06   &  0.09 & 0.03	   \\ \hline
4.12    &     271.21  &   4.46   &  0.08 & 0.05\\ \hline
4.64    &     271.33  &    4.82  &  0.07 & 0.06	   \\ \hline
5.16    &     271.40  &    5.11  &  0.06 & 0.08\\ \hline
\end{tabular}
\label{tab-pwv}
\end{table}

\subsection{PWV from tipper opacity}
Once we have produced the described fits 
for a given list of input PWV, we find a relation between the PWV and the fitted atmospheric opacity, with the aim of converting the full stream of tipper data from opacity ($\tau$) to PWV. The relation between the two quantities is described as follows:

\begin{equation}\eqnum{2}
 \tau\>=\>m\cdot PWV\>+\>c 
\label{eq2}
\end{equation}

\noindent where $m$ is the slope for the linear relation and describes the atmospheric extinction per mm of PWV, and $c$ is the dry air opacity due to molecular components.\\

\begin{figure}[h!]
\noindent\includegraphics[width=20.5pc]{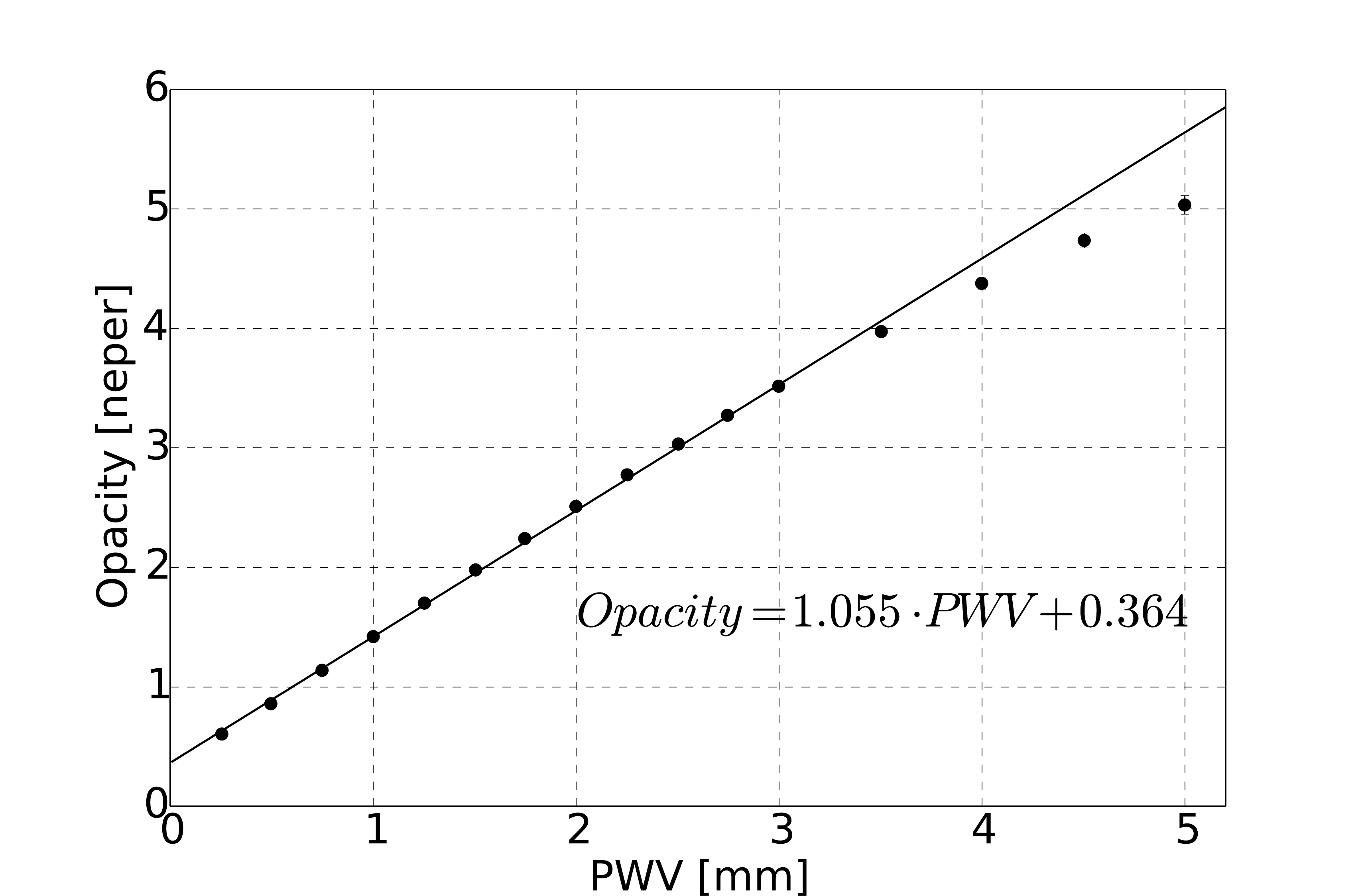}
\caption{Relationship between opacity and PWV for Chajnantor Plateau, CBI site.}
\label{fig2}
\end{figure}

\begin{figure}[h!]
\noindent\includegraphics[width=20.5pc]{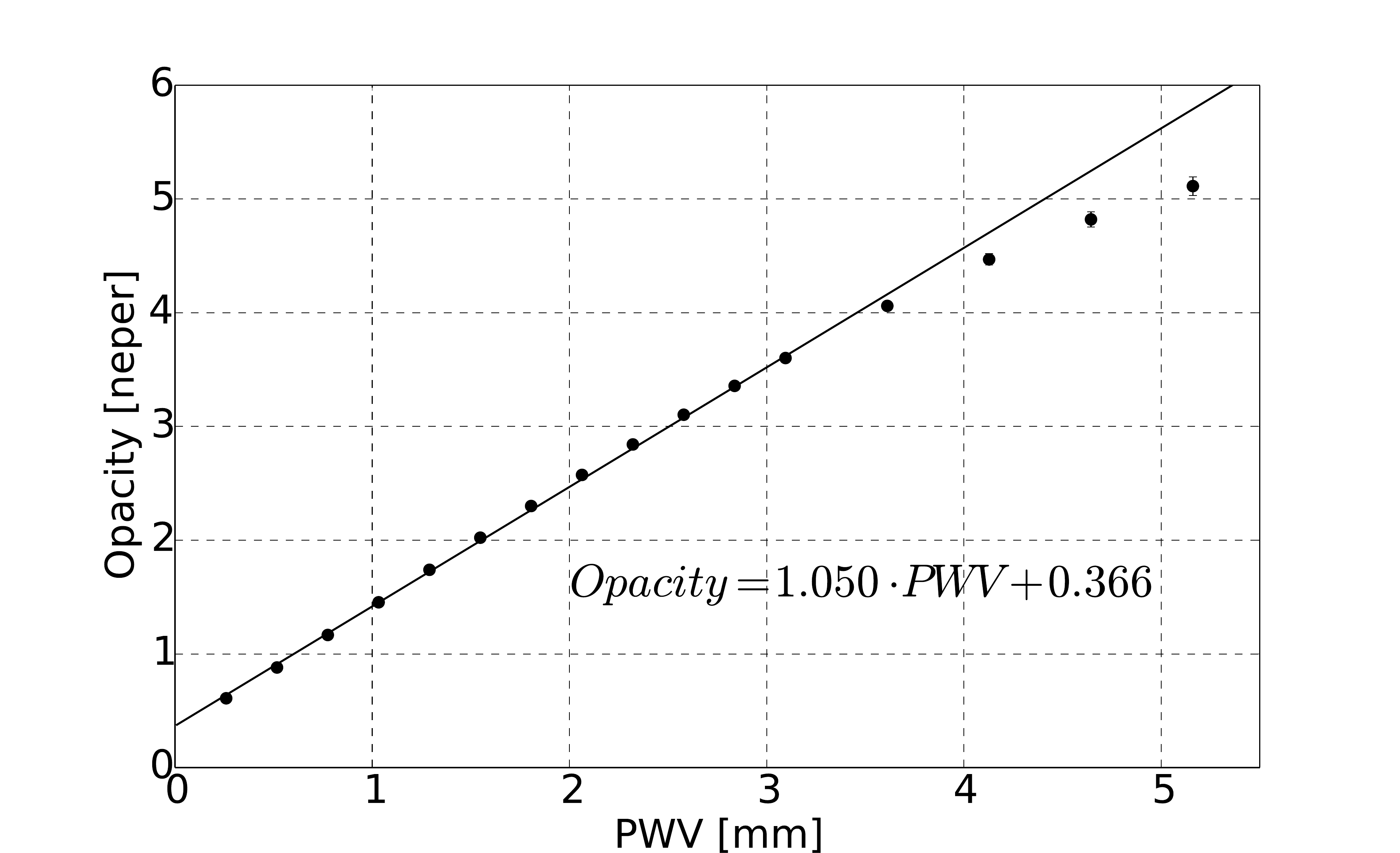}
\caption{Relationship between atmospheric opacity and PWV for Chajnantor Plateau, APEX site. Linear fit parameters are in very good agreement with \cite{radford11}.}
\label{fig3a}
\end{figure}

\begin{figure}[h!]
\noindent\includegraphics[width=20.5pc]{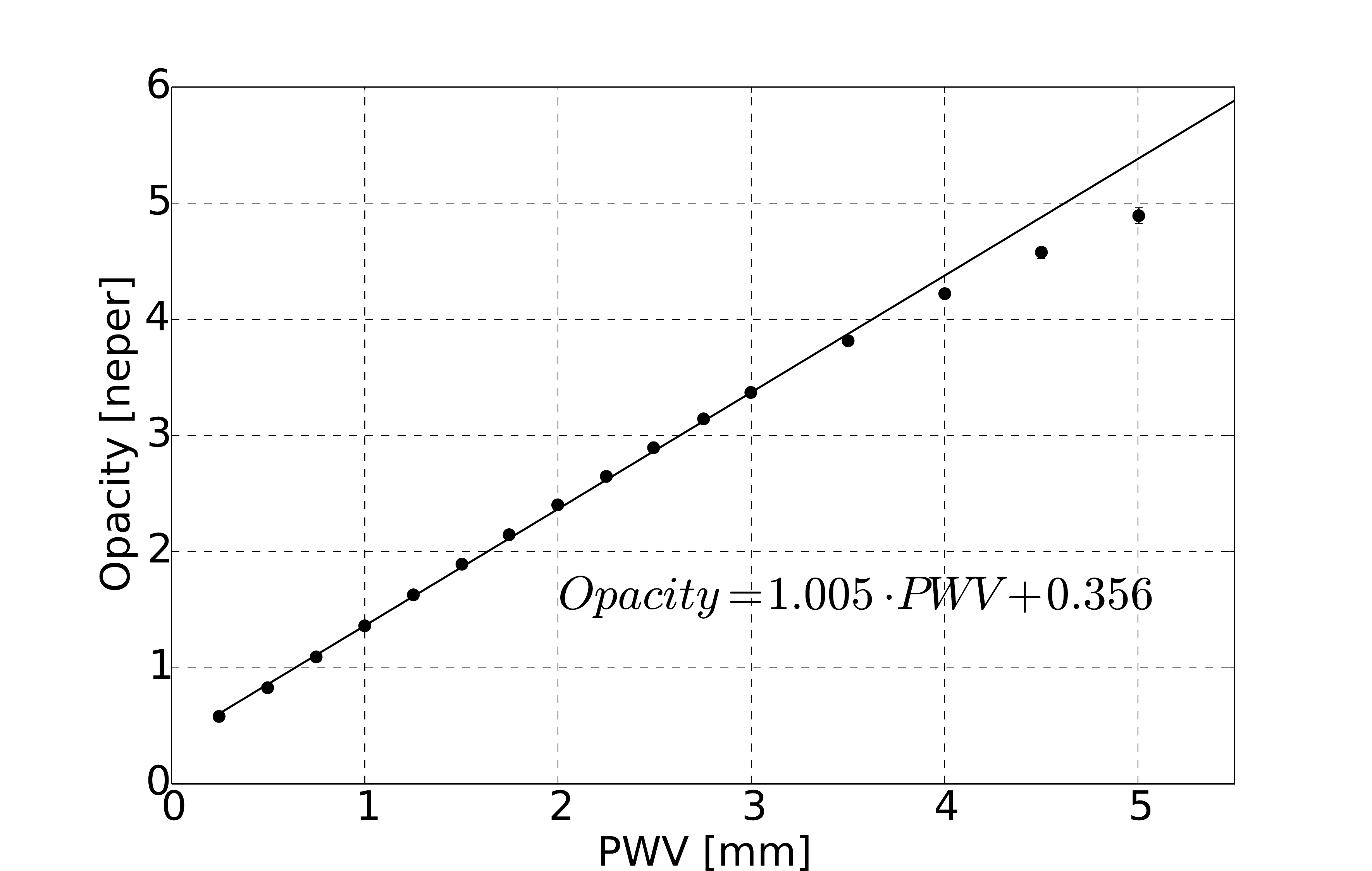}
\caption{Relationship between opacity and PWV for Cerro Chajnantor Summit.}
\label{fig3b}
\end{figure}

The fitting process to find the linear relation between PWV and opacity is applied at the 3 sites of interest: CBI site, APEX site, both at the Chajnantor Plateau, and the summit of Cerro Chajnantor. The results for the fits on these cases are shown in Figure \ref{fig2} for CBI site, Figure \ref{fig3a} for the APEX site and Figure \ref{fig3b} for Cerro Chajnantor. 

The differences between the model parameters, $m$ and $c$, for the shown $\tau$ to PWV conversions are consistent with the expected reduction of atmospheric extinction and dry air opacity with increasing altitude, as shown in Table \ref{tab_ext}. The dry air opacity derived for the CBI site was lower than the APEX site value by a small amount, nevertheless the difference is within the error of the fitting procedure. 

\begin{table}[h]
\caption{  The reduction of atmospheric extinction and opacity with increases the altitude  }
\centering
\begin{tabular}{|c |c | c | c | c |c| }
\hline
  & Altitude  & Extinction            & Dry air opacity            \\ 
     &    (m) & (neper/mm) & (neper)   \\ \hline 
 CBI site   &   5080             &   1.055        &     0.364          \\ \hline 
   APEX site &   5107             &   1.050        &     0.366          \\ \hline 
   Cerro Chajnantor  &  5612             &   1.005        &     0.356          \\ \hline 
\end{tabular}
\label{tab_ext}
\end{table}

\section{Calibration of converted PWV from sub-mm tippers}
\begin{figure}[!h]
\noindent\includegraphics[width=20pc]{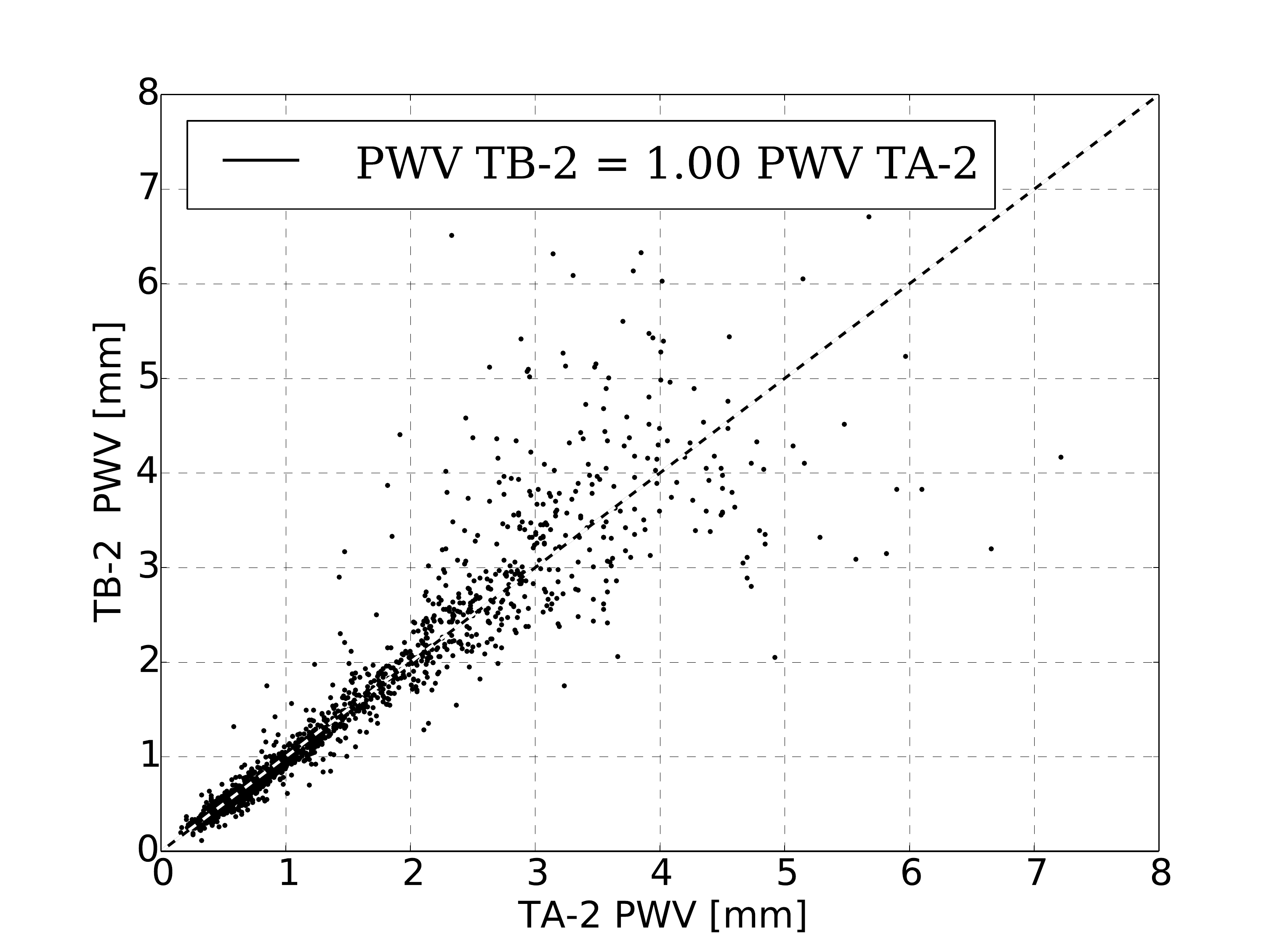}
\caption{Data from 2005-2009 are included in this figure. The dots are the time matched samples between TB-2 and TA-2. The instruments were co-located at the CBI site. The unity slope indicates a successful calibration between the instruments.}
\label{fig8}
\end{figure}

We consider the APEX PWV measurement as our calibration standard. Calibration factors for the sub-mm tippers were determined during periods were the instruments were co-located with the reference. Tipper radiometer A was installed at the APEX site, hence the converted tipper PWV values should match the APEX measurements during the period of overlap. The long-term calibration factor for Tipper A was found to be$1.088$. There was another substantial period of time in which both tipper radiometers were co-located at the CBI site. Using a calibrated datastream for Tipper A, we derive a calibration factor of $1.080$ for Tipper B, close to the one determined for Tipper A and consistent with the similar instrumental nature of both instruments. Under these conditions, the converted PWV measurements from both tippers should be the same, and this is clearly shown in Figure \ref{fig8}. 

\section{Comparison and correlations in PWV} 
Given the long-term nature of this study, we calculate relations between the sites that will establish a general difference in their atmospheric conditions, using all the data available. This would help in situations were new observatories can anticipate the overall atmospheric conditions for the area, and plan their operations and budget accordingly. The way we show the difference between the atmospheric conditions at two given sites, is by using PWV vs PWV  scatter plots with data samples previously matched and filtered in time, which was done using TOPCAT. The need for the time matching is because of the very different time sampling among all the instruments. The APEX WVR has a one-minute resolution while the tipping radiometer has a 13-minute resolution. \\

The method to compare the sites was implemented as follows. A linear regression was applied to the PWV-PWV plots using Scipy (http://www.scipy.org). The linear regression was forced to go through the origin of the cartesian system. Under this condition, the slope is the only free parameter and it represents the average difference, over several years, in zenith water vapor content between the sites. PWV measured with the radiometer at APEX was used as the reference. \\

\begin{figure}[!H]
\noindent\includegraphics[width=20pc]{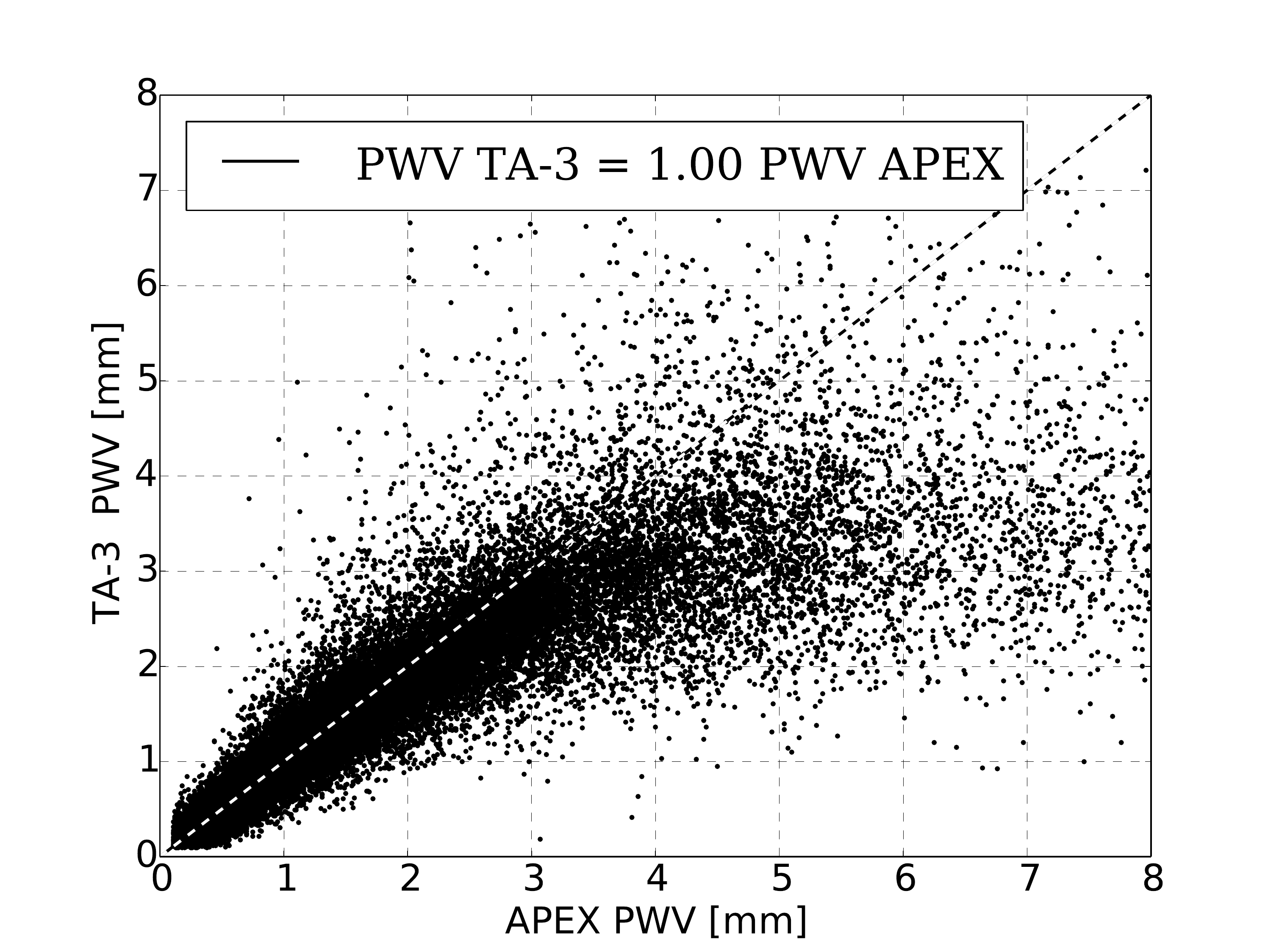}
\caption{Four years of data are shown in this figure. The dots are the time matched samples between tipper and APEX radiometer, both instruments are located in the Chajnantor Plateau. The linear regression is shown by dotted line and the numerical relation between the measurements is show in the inset.}
\label{fig6}
\end{figure}

\begin{figure}[!H]
\noindent\includegraphics[width=20pc]{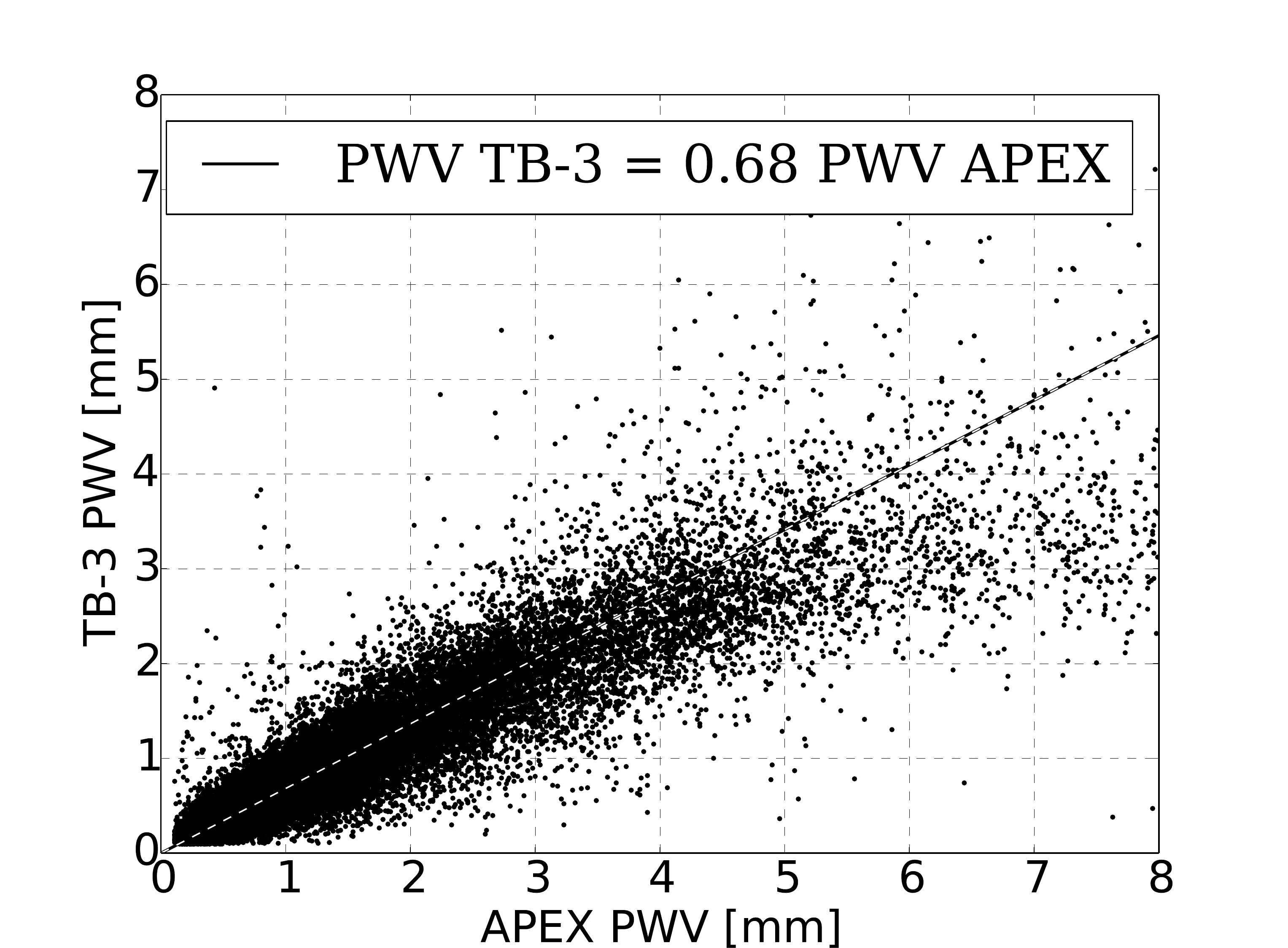}
\caption{Similar as Figure \ref{fig6}. Data from 2009-2012 are included in this figure. The lower slope in this graph shows the difference in atmospheric state between Cerro Chajnantor summit and Chajnantor Plateau.}
\label{fig7}
\end{figure}

The linear regression is only valid over a range with low PWV data due to non-linearities in the tippers. The linear range was determined by restricting the input to the fitting algorithm in $PWV_{APEX}$ ranges. The first range for $PWV_{APEX}$ was defined between $0$ and $1$ mm. Subsequent ranges increased the upper end by $0.5$ mm. A linear indicator was extracted by taking the mean of the slope between the two first ranges. Slopes for subsequent ranges were compared to the linear indicator, and the dataset increased until the calculated slope had changed by more than $3\%$ from the linear indicator. With these method, the data included in the final fits were always over $70\%$ of the available data. This laborious procedure to calculate the slope was necessary due to the non-linearity exhibited by the sub-millimeter tippers. Compression in the tipper radiometric performance give rise to bent tails for high PWV values shown in Figures \ref{fig6} and \ref{fig7}. \\

As an example for the results of this analysis, we show a time matched correlation plot for APEX vs TA-3, both co-located at the APEX site, Fig. \ref{fig6}. Results for the analysis between Cerro Chajnantor and the plateau are shown in Fig. \ref{fig7}, where a time matched correlation plot for APEX vs TB-3 is presented. These plots show evidence of the overall lower PWV at Cerro Chajnantor compared to the Plateau because of the altitude difference between the sites. We acknowledge the existence of inversion layers appearing in certain periods, in which the PWV at Cerro Chajnantor lowers dramatically compared to the PWV at the Plateau, as shown in Figure~\ref{fig4} and also described in \cite{bustos14}. The detailed short-term study to reveal this effect in depth is out of the scope of this paper and will be addressed in a future publication.

The results for the PWV slopes found in this study, using all the instruments and sites are shown in Table \ref{tabla-instr}.

\begin{table}[h]
\caption{ Summary with the results of PWV slopes between different sites, time span of measurements and altitude difference  between sites}
\centering
\begin{tabular}{|c |c | c | c | c |c| }
\hline
  Instruments                    &     Period    of years    &     Slope       &    Altitude   \\ 
   pairs               &                                  & & Difference \\
                 &                                  & &  (m) \\ \hline \hline
 {\bf{ TA-2 / APEX}}   &  {{ 2006-2010 }}        &  {\bf{ 1.07 }}       &   -27 \\ \hline\hline
   {\bf{ TB-2 / TA-2 }}        &  {{  2005-2009 }}         &  {\bf{  1.00   }} &  0 \\ \hline\hline
 {\bf{ TA-3 / APEX}}                    &  {{  2011-2014 }}        &  {\bf{ 1.00 }}   &   0    \\ \hline \hline
 {\bf{ TOCO / APEX }}            &    {{2009}}        &   {\bf{0.89}}     &   213  \\ \hline \hline
 {\bf{TB-3 / APEX }}    &     {{2009-2012}}      &       {\bf{0.68}}     &  505\\ \hline\hline
 {\bf{ TB-3 / TA-2}}                     &  {{ 2009-2010}}        &  {\bf{0.68}}   &  532   \\ 
\hline                  
\end{tabular}
\label{tabla-instr}
\end{table}

The TOCO/APEX PWV ratio is higher than one would expect from the altitude difference of the sites. We noted that this result might be biased from the fact that the RHUBC-II measurements were only taken during daytime, when the atmosphere is more mixed and the appearance of inversion layers are improbable, as opposed to all the other measurements which are taken irrespective of hour of the day. More data would be needed for this site to clarify the difference and draw conclusions.\\

The TAO data and TAO/APEX PWV ratios were not shown in this work because the amount of data was insufficient for the analysis. Besides that, we noted that the PWV measured at TAO and TB-3, were in very good agreement as expected.\\

For all instruments pairs in Table \ref{tabla-instr} (except for TOCO/APEX), the results are consistent with the exponential decay in PWV of a standard atmosphere. This is represented as follows:

\begin{equation}\eqnum{5}
\displaystyle PWV\>=\> PWV_{0}\cdot {\rm e}^{-\frac{\Delta h}{h_{o}}}
\label{eq1}
\end{equation}

\noindent where $PWV_{0}$ is the PWV measured at the lowest level, $h_{0}$ the scale height, and $\Delta h$ the altitude difference between the sites. We calculated the atmospheric scale height for the relevant sites and with the information included in Table \ref{tabla-instr}. The results are summarized in Table \ref{tabla-instr2}.

\begin{table}[h]
\caption{ Atmospheric scale heights calculated from the PWV ratios and altitude differences shown above.}
\centering
\begin{tabular}{|c |c | c | c | c |c| }
\hline
  Involved sites               &              Scale Height             \\ \hline  \hline  

APEX site - Cerro Chajnantor             &       1309          (m)                    \\
                                                            &      \\ 
CBI site - Cerro Chajnantor             &       1379          (m)                    \\
\hline                  
\end{tabular}
\label{tabla-instr2}
\end{table}

The scale heights reported in this paper, Table \ref{tabla-instr2}, are within the suggested values shown in \cite{radford16}.

\section{Validation of the method}
\begin{figure*}[!H]
\noindent\includegraphics[width=30pc]{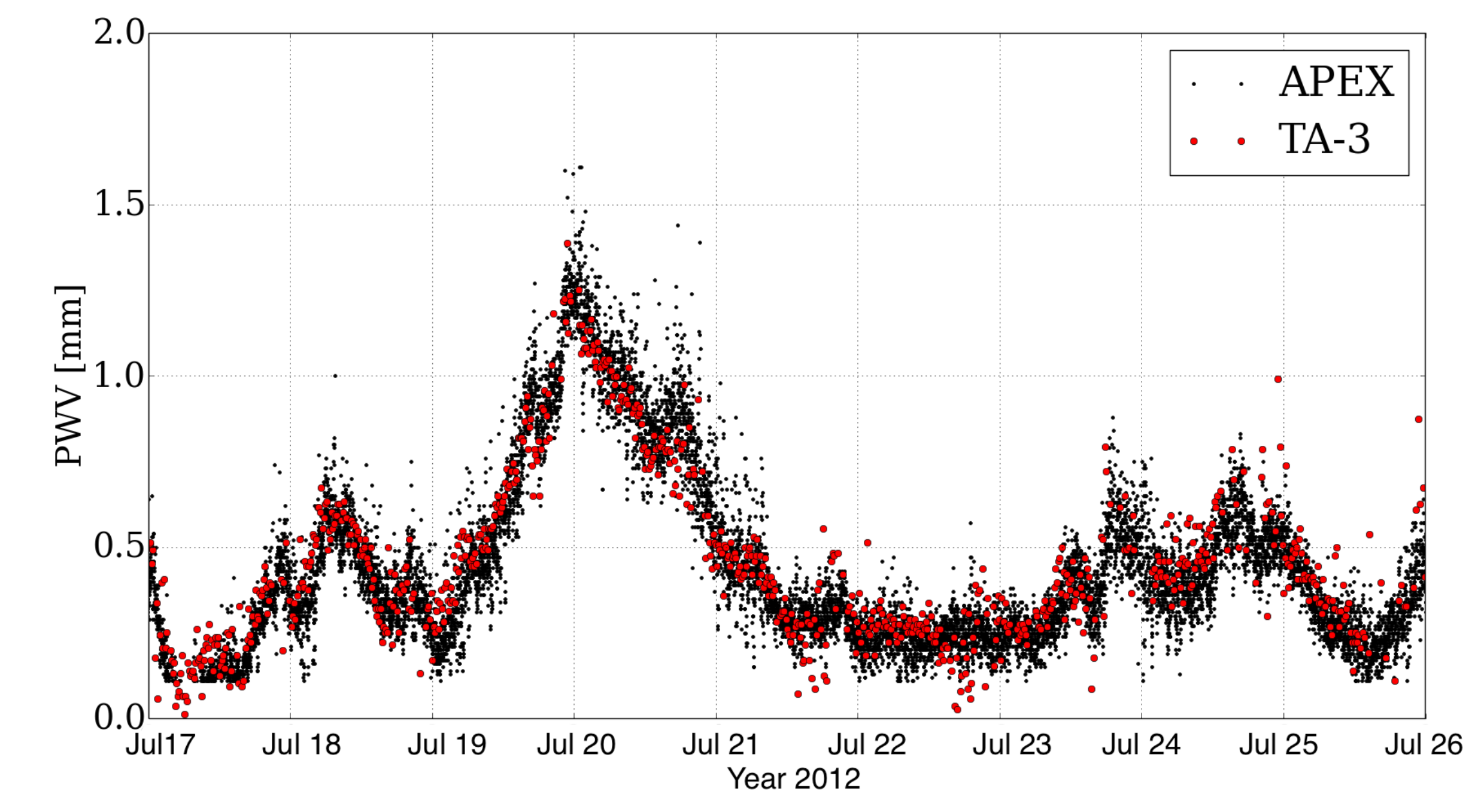}
\caption{PWV converted data from TA-3 compared with the 9-day analysis between PWV from APEX. TA-3 converted PWV agrees very well with APEX since they are only 4 meters apart, validating the conversion method.}
\label{fig4b}
\end{figure*}

\begin{figure*}[!htbp]
\noindent\includegraphics[width=39pc]{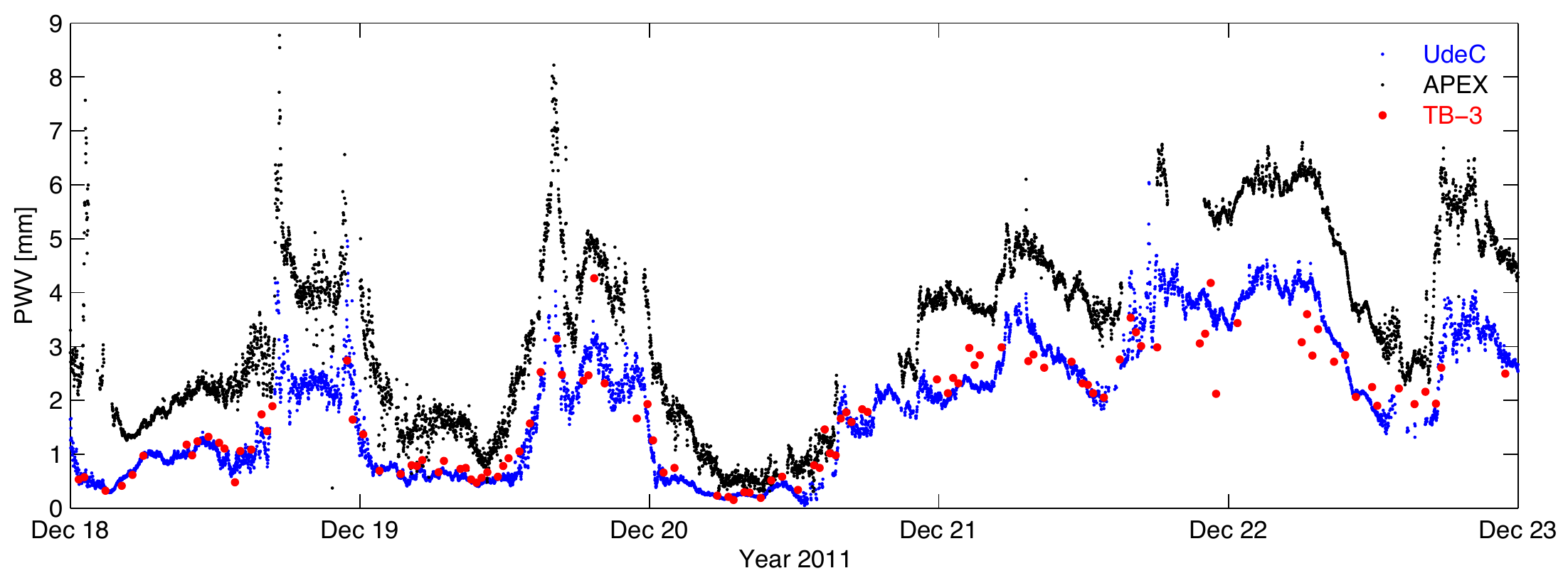}
\caption{PWV converted data from TB-3 compared with the 5-day analysis between PWV from APEX and UdeC, shown by \cite{bustos14}. TB-3 converted PWV agrees very well with UdeC since they are only 5 meters apart, validating the conversion method.}
\label{fig4}
\end{figure*}

We validated our transformation model and calibration by comparing the results with the PWV from APEX taken concurrently, as shown in Figure \ref{fig4b}, and with a totally independent reference as is the UdeC 183 GHz WVR, shown in Figure \ref{fig4}. The UdeC and TB-3 instruments are located only 5 meters apart, at the CCAT weather monitoring station.

The PWV estimated from the TB-3 tipper radiometer (red dots in Figure \ref{fig4}) are in very good agreement with the UdeC data (blue dots in Figure \ref{fig4}). A noticeable PWV difference  between TB-3 and the APEX radiometer is seen due to the altitude difference between the sites, as it was also found and explained in \cite{bustos14}. Discrepancies seen around December 22nd between UdeC and TB-3 are due to non-linearity in TB-3. The non-linearity effect is not appreciable in Figure \ref{fig4b} because the PWV during the period was low, hence, we suggest that the tipper data can only be used truthfully up to $3$ mm of PWV.


\newpage

\section{Conclusions}

This work collected $> 3\cdot 10^{6}$ data coming from different instruments and locations over the Chajnantor area and for long periods of time. 

The AM v9.0 has been recently updated and the use of the modified version prove crucial to get the results shown here. The method used in this work to convert atmospheric opacity from the sub-mm tippers to PWV, has been validated by the instruments at these sites and turned out to be a useful tool for atmospheric studies in the Chajnantor area.


The tipper converted PWV data was calibrated using APEX PWV measurements.
 
When co-located at the Plateau, and once the data was properly calibrated, both tipper radiometers measured the same amount of PWV. This fact allowed us to use the tipper measurements when the instruments were taken to different locations in the area. 

The CBI site shows 7\% excess of PWV compared to the APEX site. This difference is due to the altitude difference between the sites. 

For Cerro Chajnantor summit, the PWV ratio with the Plateau is 0.68, which indicates 32\% less PWV compared to the Plateau and for our long term study. Although this value is close to previous results, i.e. cite{bustos14} and \cite{radford16}, our study is longer term and therefore averages out temporal variabilities in the atmospheric quantities.

Cerro Toco is a interesting site to study the amount of PWV since it is located between the Chajnantor Plateau and Cerro Chajnantor summit. Our result for PWV ratio between Toco and the Plateau is subject to selection bias, as explained in the text. More data for this site would be needed to estimate and clarify the observed PWV ratios between Toco and APEX.

For the interested reader, the AM configuration files used in this paper can be requested by e-mail to F. Cort\'es (fercortes@udec.cl).





%
%
%
%
%
%
%

\begin{acknowledgments}\\

We thank the people and observatories who contributed to this project. Simon Radford to provide us data of tipping radiometers and helped to understand the variables. Eli Mlawer for the Cerro Toco summit data. Kentaro Motohara for TAO PWV data. We very much appreciate the contribution by Scott Paine, who provided us deep insight on the use of the AM software.

This publication is based on data acquired with the Atacama Pathfinder Experiment (APEX). APEX is a collaboration between the Max-Planck-Institut fur Radioastronomie, the European Southern Observatory, and the Onsala Space Observatory.

R. Reeves and F. Cort\'es acknowledge  support from Fondo Gemini-Conicyt programa de astronom\'ia/pci folio $32140030$. R. Reeves also thank support by CATA Basal PFB-06 Etapa II.

\end{acknowledgments}

\end{article}
%
%
%
%
%
%
%
%


\end{document}